\documentclass[reprint,twocolumn,prb,amsmath,amssymb,article,showpacs,aps]{revtex4-2}
\usepackage{graphicx}
\usepackage{dcolumn}
\usepackage{blindtext}
\usepackage{bm}
\usepackage{tikz}
\usepackage{pgfplots}
\pgfplotsset{compat=1.14}
\usepackage{adjustbox}
\usetikzlibrary{decorations.pathreplacing}
\usepackage{amssymb,float}
\usepackage{float}

\usepackage{bm}
\usepackage{multirow}
\usepackage[colorlinks,plainpages=false,linkcolor=red,urlcolor=blue,citecolor=blue,filecolor=black,pdfpagemode=UseNone]{hyperref}

\usepackage{adjustbox}

\usepackage{amsmath}
\usepackage{lipsum}
\usepackage{rotating}
\usepackage{color}
\usepackage{xcolor}
\usepackage{pgfplots}
\usepackage{epsfig}
\usepackage{epstopdf}
\usepackage{bm}
\usepackage{textcomp}
\usepackage{array}
\usepackage{upgreek}
\usepackage{ctable}
\usepackage{xfrac}
\usepackage{mathtools}
\usepackage{amsfonts}
\usepackage{booktabs}
\usepackage{lineno}      
\usepackage{cleveref}    

\begin{document}

\title{First-principles study of optoelectronic and thermoelectronic properties \\of the ScAgC half-Heusler compound}
\author{Vinod Kumar Solet$^{1,}$}
\altaffiliation{vsolet5@gmail.com}
\author{Shamim Sk$^{2}$}
\author{Sudhir K. Pandey$^{1,}$}
\altaffiliation{sudhir@iitmandi.ac.in}
\affiliation{$^{1}$School of Mechanical and Materials Engineering, Indian Institute of Technology Mandi, Kamand - 175075, India}
\affiliation{$^{2}$School of Physical Sciences, Indian Institute of Technology Mandi, Kamand – 175075, India}

\begin{abstract}
\setlength{\parindent}{4em}

In this paper, we have presented a theoretical study in the context of photovoltaic (PV) and thermoelectric (TE) applications of ScAgC. The electronic, optical, and thermoelectric properties have been investigated systematically using DFT and semi-classical Boltzmann transport theory. DFT calculates a direct band-gap of $\sim$$0.47$ eV, whereas the $G_{0}W_{0}$ method estimates a band-gap of $\sim$1.01 eV. We used parabola fitting to estimate the effective mass ($m^{*}$) values for bands B1–B4 at $\Gamma$-point, which are $\sim$ -0.087 (-0.075), $\sim$ -0.17 (-0.27), $\sim$ -0.17 (-0.27), and $\sim$ 0.049 (0.058) along the $\Gamma$-$X$ ($\Gamma$-$L$) direction, respectively. We have investigated phonon dispersion and thermal properties. Furthermore, the properties of optoelectronics are calculated and analysed over a range of photon energies from 0 to 10 eV. The optical conductivity $\sigma (\omega)$, refractive index $\tilde{n}(\omega)$, and dielectric function $\epsilon (\omega)$ show strong optical transitions in the visible region. The lowest calculated value of reflectivity r($\omega$) is $\sim$0.24 at $\sim$4.7 eV, and the highest calculated value of absorption coefficient $\alpha (\omega )$ is $\sim$$1.7\times 10^{6}$ $cm^{-1}$ at $\sim$8.5 eV. At 300 K, we have expected a maximum solar efficiency (SLME) of $\sim$33\% at $\sim$1 $\mu m$ of thickness. The lattice part of thermal conductivity $\kappa_{ph}$ shows a maximum value of $\sim$3.8 W$m^{-1}K^{-1}$ at 1200 K. At 1200 K, for electron doping of $\sim$3.9$\times$$10^{21}$$cm^{-3}$, the maximum value of $S^{2}\sigma /\tau$ is $\sim$145 $\times 10^{14}$ $\mu WK^{-2}cm^{-1}s^{-1}$, while for hole doping of $\sim$1.5$\times$$10^{21}$$cm^{-3}$, it is $\sim$123 $\times 10^{14}$ $\mu WK^{-2}cm^{-1}s^{-1}$. The highest $ZT$ at 1200 K is expected to be $\sim$0.53, whereas the optimal \%efficiency is predicted to be $\sim$8.5\% for cold and hot temperatures of 300 K and 1200 K, respectively. The collected results suggest that the ScAgC compound would be a potential candidate for renewable energy sources such as solar cell and TE applications.
      
\end{abstract}

\maketitle

\section{Introduction} 
\setlength{\parindent}{3em}
Energy is so crucial in everyday life because it is a basic human requirement. Energy harvesting is gaining importance as a way to recover and use any free energy that is available from the environment or waste from industrial systems. Fossil fuels (oil, coal, and natural gas) are a more convenient way to generate electricity \cite{conti2016}. But, burning of fossil-fuel is the serious concern for human health and the future, as well as being a significant contributor to global energy disasters and environmental degradation. Also, fossil fuels are not renewable sources of energy since they are limited. If we want to solve these challenges, then we need to use renewable energy sources such as solar photovoltaic \cite{liu2013,nelson2003}, thermoelectric generators (TEGs) \cite{disalvo1999}, bio-mass, hydro-power, and wind-energy. It is believed that these renewable energy sources will hold the key to fulfill the future's energy demand. Among them, PV solar cells are one of the safest renewable resources since they produce the least amount of pollution. PV solar cells are composed of semiconductor materials that become conductive when exposed to light. In the same way, TEGs, which are made up of TE materials, have a lot of potential to help get around these problems because they can transform unwanted waste heat energy generated by a heat source into electricity. The performance of a device for commercial use can be expressed by its efficiency. Thus, there have been ongoing efforts in these fields of research to improve efficiency as well as to find novel materials that can function as PV and TEGs.

Developing new materials for PV and TE applications is a crucial part of these fields. A lot of semiconducting materials, such as complex chalcogenides \cite{khare2020,chen2003}, hybrid perovskites \cite{zhang2016}, skutterudites \cite{chen2003}, and Heusler compounds \cite{chen2003}, have been examined as potential PV and TE devices. Among these materials, Heusler compounds have shown a lot of interest to researchers \cite{zhu2018,zhu2019} since the first discovery of the Heusler alloy by the German scientist Friedrich Heusler in the early 1900s \cite{heusler1903}. Heusler alloys are an impersive class of compounds with great potential for use in different energy applications, including optoelectronics \cite{yu2012SLME,mehnane2012,gruhn2010}, TE \cite{graf2011,sharma2014JPCM}, piezoelectric \cite{roy2012}, spintronics \cite{graf2011,ma2017}, and topological insulators \cite{graf2011,pandey2021}, $etc$. Heusler compounds are classified into three categories depending on their simplest crystalline structure and chemical composition. The first one is the full-Heusler alloys ($X_{2}YZ$), which have a space group of Fm$\bar{3}$m (No. 225) \cite{shastri2019,sharma2014}. A quaternary Heusler compound is formed when one of the $X$ atom is replaced by another $X^{'}$ atom, \textit{i.e.} $XX^{'}YZ$ \cite{gencer2021}. When one of the sublattices is vacant, \textit{i.e.}, in an $XYZ$ composition, the materials are referred to as half-Heusler (HH) compounds. Both quaternary Heusler and HH compounds crystallize in cubic phase with space group of F$\bar{4}$3m (No. 216) \cite{chen2003,gencer2021}. Basically, $X$, $X^{'}$ and $Y$ are transition elements, while $Z$ belongs to the $p$-block element. Moreover, the number of transition metal elements and their valence electrons affect the physical properties of these compounds. Compounds with 8-valence electrons show a stable ground state structure \cite{vikram2019}. The fundamental physical properties of the 8-valence electron compounds are comparable to those of standard semiconductors \cite{shrivastava2019}.

In 1983, De Groot \textit{et al.} \cite{de1983} performed the first computational study of HH compounds. Following that, several HH alloys have been discovered to be semiconductors in various theoretical and experimental contexts \cite{surucu2019,shastri_ZrNiSn}. HH compounds have a variable band-gap (0 to 4 eV), which makes them good for use in solar cells and TE applications \cite{zhu2018,casper2012}. Currently, most of the HH alloys show their application in the field of PV solar cell, because of many features related to optical properties, such as broad absorption spectra, low reflectivity, large refractive index, and so on \cite{amudhavalli2018,touia2021}. In the field of TE, these compounds have a high power factor ($PF$), a high figure of merit ($ZT$) value, a high density of states in the vicinity of Fermi level, and the presence of a flat band structure \cite{zhu2019,fu2015}. An efficient TE material would have a high $PF$ and a low thermal conductivity. However, for the wide use of power production of TE materials, an enormous improvement in conversion efficiency is needed, which is basically decided by $ZT$. Thomas Gruhn \cite{gruhn2010} has studied various types of HH compounds for optoelectronic properties that are based on DFT calculations. Of these, only six compounds meet the proper requirements for optoelectronic properties like PV applications. In the search for PV solar cell and TE applications, we found that ScAgC could be a good candidate for future energy-conversion. ScAgC comes in the class of I-III-IV alloys within the family of 8-valence electron HH compounds. The electronic structure and related properties of this compound have been reported previously \cite{belmiloud2016}. But, as per our knowledge from the literature survey, this compound has not been explored for PV and TE applications. 

The goal of this study is to analyse the electronic, optical, thermal, and TE properties of ScAgC using DFT and semiclassical-Boltzmann theory. The electronic structure calculation via DFT shows the semiconducting ground state of this compound with an energy gap of $\sim$ 0.47 eV. The band-gap from $G_{0}W_{0}$ method is calculated to be $\sim$ 1.01 eV. Effective mass values are also estimated at the band edge curvature. Phonon dispersion and the lattice contributions to thermodynamic properties such as Helmholtz free energy, internal energy, entropy, and constant-volume specific heat have been investigated. The Debye temperature $\Theta_{D}$ is obtained as $\sim$ 629 K. We estimate the optical properties with respect to different energy regions in order to further evaluate the compound for PV applications. The maximum absorption is predicted to be $\sim 1.7 \times 10^{6}$ $cm^{-1}$ at $\sim$ 8.5 eV, while the minimum reflectivity is discovered to be $\sim$ 0.24 at $\sim$ 4.7 eV. We have predicted a maximum spectroscopic limited maximum efficiency (SLME) of $\sim$ 33\% at around 1 $\mu m$ thickness. To understand the material's transport behaviour, $PF$ per relaxation time is calculated and shows a maximum value of $\sim$ 145 (123) $\times 10^{14}$ $\mu WK^{-2}cm^{-1}s^{-1}$ at 1200 K for electron (hole) doping compound. After that, the $\kappa_{ph}$ values are computed, yielding a maximum value of $\sim$ 3.8 W$m^{-1}K^{-1}$ at 1200 K. To investigate the material further for TE applications, we calculated the temperature-dependent $ZT$ and $\eta$ (\%efficiency) for ScAgC with the highest $PF$, yielding electron and hole doping concentrations by taking into account the two values of electronic relaxation time ($\tau$). For this compound, maximum $ZT$ of $\sim$ 0.53 and $\sim$ 0.47 are estimated for $p$-type and $n$-type, respectively. The maximum $\eta$ of $\sim$ 8.5\% for the $p$-type and $\sim$ 7.4\% for the $n$-type is expected for a cold side temperature of 300 K and a hot side temperature of 1200 K. The calculated values of the lattice parameter and band-gap are comparable to other solar materials like Si \cite{barla1984,nelson2003}. As a result, the studied parameters suggest that this compound can be used as a promising candidate for sustainable energy sources, such as TE and solar cell applications.

\section{Computational details}
The projected augmented wave (PAW) method \cite{PAW} has been used to calculate the electronic and optical properties of the ScAgC compound in the ABINIT \cite{abinit} software. All the calculations have been done in the framework of first-principles DFT, in which the Kohn-Sham (KS) equation has been solved. The Perdew, Burke, and Ernzerhof (PBE) functional with a generalized-gradient approximation (GGA) is used to solve the exchange and correlation (XC) potential in the KS equation \cite{GGA-PBE}. The Monkhorst-Pack grids \cite{monkhorst} as 16x16x16 $\textbf{k}$-points mesh is used to calculate electronic structure, while an unshifted 6x6x6 $\textbf{k}$-points grid is used to compute band-gap using the $G_{0}W_{0}$ method. For the plane wave, a kinetic energy cutoff (\textit{ecut}) of 25 Ha is used, and a two-times larger \textit{ecut} value is used for the PAW energy cutoff to achieve good total energy convergence in PAW computations. The lattice parameter is taken from the literature \cite{belmiloud2016}, and this is used in lattice parameter optimization. The lattice parameter is optimized and we also checked the convergence with respect to \textit{ecut}. The obtained value of the optimized lattice parameter is 5.600 \AA (10.5825 Bohr). Using this converged lattice parameter and the corresponding \textit{ecut}, all the \textit{ab-initio} calculations have been done. 

To perform linear-response calculations, we used the density functional perturbation theory (DFPT) with a four-shifted $\textbf{q}$-points mesh of 4x4x4 \cite{gonze1997first,gonze1997dynamical}. We got a dynamical matrix, and the eigenvalues of this dynamical matrix were used to figure out the phonon frequencies. A large $\textbf{k}$-point mesh of 50x50x50 is used to obtain the necessary matrix elements for electrons in order to further calculation of optical properties. These properties are estimated using Kubo–Greenwood formulation \cite{mazevet2010} in the conducti utility \cite{conducti}. We have used a Gaussian width of 150 meV in the conducti calculation. The theoretical tool SLME in the JARVIS package \cite{yu2012SLME} is used to estimate the efficiency of this compound. A heavy $\textbf{k}$-points mesh of 30x30x30 was used to generate energy files in ABINIT software for further calculation of transport properties. These properties are obtained by interpolating the DFT band structure in the BoltzTraP code \cite{madsen2006}. In this code, semi-classical Boltzmann transport theory within constant relaxation time approximation has been used. 

ScAgC HH compound has a $C1_{b}$ structure with F$\bar{4}$3m (group no. 216) space group. In this FCC structure, atoms are in an equal composition of 1:1:1, having a Wyckoff position \cite{hahn1983} of Sc is 4b$(0.5,0.5,0.5)$, Ag is 4a$(0,0,0)$, and C is 4c$(0.25,0.25,0.25)$ \cite{gruhn2010}, in units of a cubic lattice constant. The 4$s$, 3$p$, and 3$d$ electrons have been considered valence electrons for the Scandium atom. In the Silver atom, the 5$s$ and 4$d$ electrons are taken for valence electrons, while for Carbon, 2$s$ and 2$p$ are valence states.

\section{Results and discussion}

\subsection{\label{sec:level2}Electronic properties}
The knowledge of electronic band structure is required for the study of PV and TE applications of the materials. The aim of this section is to determine the electronic band-gap and analyze the electronic band structure because the optical and TE properties are highly dependent on the band structure of the materials. We have presented on Fig. 1 the dispersion curve of ScAgC along the high symmetry $\textbf{k}$-path $W$-$L$-$\Gamma$-$X$-$W$-$K$ within the first Brillioun zone. The calculation was done in the energy range of -5 to 5 eV. It is evident that ScAgC shows semiconducting behaviour with a direct band-gap of $\sim$ 0.47 eV, where the valence band maximum (VBM) and conduction band minimum (CBM) occur at a high symmetry $\Gamma$-point. The ability to use direct band-gap compounds in PV solar cells is one of their distinguishing characteristics \cite{shaposhnikov2012}. The horizontal dashed line represents the Fermi level ($E_{F}$), which is set at the middle of the band-gap. At higher temperatures, direct transitions play a significant role in the optical and transport properties of materials. Band 1, band 2, and band 3 are degenerate at $\Gamma$-point. The degeneracy is not lifted along $L$ to the $X$-direction for the bands 2 and 3, and it becomes non-degenerate at the $W$-point.  

It is critical to estimate effective masses ($m^{*}$) as accurately and precisely as possible when describing features such as transport, optoelectronic, thermoelectric, and so on \cite{ashcroft1976}. In particular, $m^{*}$ plays a crucial role in the design of solar cells even though the compound is extremely absorbent since heavy charge carriers indicate poor efficiency. Along the high symmetric direction, the effective mass of the charge carriers (electrons and holes) in terms of the free electron mass ($m$) at the $\Gamma$-point has been calculated. In Fig. 1, the bands that make a significant contribution to the transport and optical properties of the ScAgC compound are marked with numbers. The symbols B1, B2, B3, and B4 represent the bands 1, 2, 3, and 4, respectively. In the Table I, $\Gamma-\Gamma$$X$ ($\Gamma-\Gamma$$L$) indicates the value of $m^{*}$ at a $\Gamma$-point along the $\Gamma$-$X$ ($\Gamma$-$L$) direction. The value of $m^{*}$ is defined by the formula $m^{*}$ = $\hbar^{2}/({\frac{d^{2}E}{dk^{2}}})$ under parabolic approximation \cite{ashcroft1976}. As per this formula, the value of $m^{*}$ at $\textbf{k}$-point is determined by the curvature of the dispersion curve. The parabola equation was applied to obtain the $m^{*}$ at the $\Gamma$-point via fitting the curve near the VBM and CBM. The obtained values of $m^{*}$ along the $\Gamma$-$X$ ($\Gamma$-$L$) direction for the bands B1-B4 are $\sim$ -0.087 (-0.075), $\sim$ -0.17 (-0.27), $\sim$ -0.17 (-0.27), and $\sim$ 0.049 (0.058), respectively. The bands B2 and B3 have the same $m^{*}$ because they are degenerate along the $L$-$X$ direction. As per results, the values of $m^{*}$ for bands 2 and 3 is larger as compared to bands 1 and 4 because the curvature is less for the degenerate bands as compared to bands 1 and 4. $m^{*}$ is related to the effect of all the internal forces on the motion of electrons or holes in their respective bands. $m^{*}$ values of ScAgC are significantly lower than those of most semiconductors \cite{janssen2016}. The lower $m^{*}$ means that when charge carriers are subjected to the potential, they can result in greater mobility. Therefore, the lighter $m^{*}$ of the charge carriers makes them ideal for high performance electronic device applications \cite{shi2013}. 

\begin{table*}
\caption{\small{Effective mass ($m^{*}$) of charge carriers calculated for ScAgC at $\Gamma$-point in various bands. }}
\resizebox{0.57\textwidth}{!}{%
\setlength{\tabcolsep}{6pt}
\begin{tabular}{@{\extracolsep{\fill}}c c c c c c c c c c c} 
\hline\hline
 
\multicolumn{1}{c}{$m^{*}$ at } & & & &  \multicolumn{4}{c}{Bands}  &&  \multicolumn{1}{c}{} & \multicolumn{1}{c}{}\\ 
\cline{3-11}                                
  $\Gamma-point$ & & \multicolumn{1}{c}{B1} &&   \multicolumn{1}{c}{B2}  &&   \multicolumn{1}{c}{B3}   &&   \multicolumn{1}{c}{B4}\\
     
 \hline
$\Gamma-\Gamma X$ && -0.087 && -0.17 && -0.17 && 0.049  \\
$\Gamma-\Gamma L$ && -0.075 && -0.27 && -0.27 && 0.058  \\

\hline\hline 
\end{tabular}}
\end{table*}

In DFT calculation, we have computed the KS band structure of ScAgC. Note that the KS DFT tends to underestimate the band-gap of the materials, as studied in previous literature \cite{yakovkin2007}. Consequently, it is necessary to go beyond DFT to get a more accurate band-gap. Then the single-shot $GW$ ($G_{0}W_{0}$) calculations are used to obtain better band-gap values because of their reasonable accuracy with experimental data, which has been proven for a wide range of materials \cite {schena2013,godby1988,zhu1991,sihi2021,morales2017,hedin1965}. The band-gap calculated with $G_{0}W_{0}$ method is $\sim$ 1.01 eV. Unfortunately, no experimental information is available to compare the electronic properties of ScAgC, while the gap obtained is in good agreement with the previous theoretical study on this compound \cite{belmiloud2016}.    
\begin{figure}
\includegraphics[width=6cm, height=6cm]{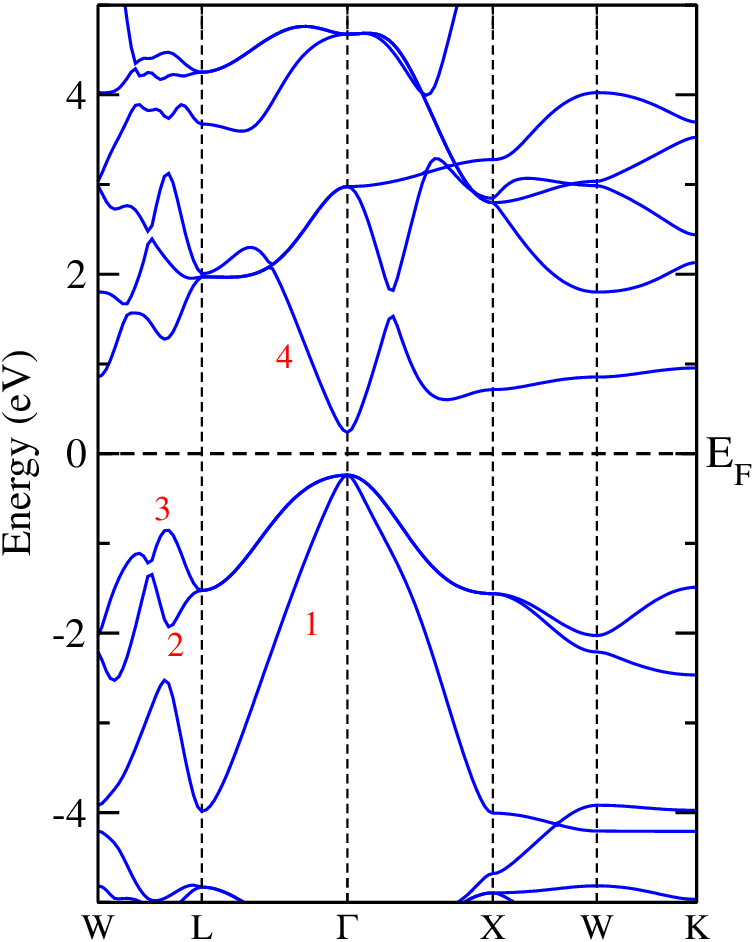} 
\caption{ Electronic dispersion curve of the ScAgC compound along the high-symmetric direction.} 
\end{figure} 

The total density of states (TDOS) and the orbital-resolved partial density of states (PDOS) are also calculated to understand the contribution of different atoms to the optical and TE properties of the material. It is clear from the TDOS, shown in Fig. 2(a), that ScAgC is a semiconducting material as predicted by the band-structure calculation. In the VB, there are mainly three peaks of A, B, and D between energies of -4 and 0 eV. These respective peaks are located at $\sim$ -3.9 eV, $\sim$ -2.4 eV, and $\sim$ -1.5 eV. The CB also contains some intense peaks from 0 to 3 eV, which will have an important aspect in the study of optical properties. The peak E is located at $\sim$ 0.9 eV, while peaks F and G are respectively situated at $\sim$ 1.8 eV and $\sim$ 2.8 eV. Figure 2(b) depicts the contribution of various states in VB and CB, as well as a band-gap near the Fermi energy ($E_{F}$). Because the Sc-3$d$, Ag-4$d$, and C-2$p$ states are mostly found near the $E_{F}$, the electric dipole transition $p$-$d$ and $d$-$d$ should contribute to the strong absorption of the ScAgC compound. A high contribution from the Sc-3$d$ states appears in CB. On the other hand, the 4$s$ and 3$p$ orbitals of Sc atom have a slight influence, and also, the $s$-orbitals of the Ag and C atoms have a minor role in deciding the properties of a material. The $d$-orbital of the Ag atom contributes significantly to the lower energy part of the VB, with some help from the C-2$p$ and Ag-5$s$ states. Energy states near the $E_{F}$ have vital role in deciding the optical and TE properties of the ScAgC compound.

\begin{figure}
\includegraphics[width=7cm, height=7cm]{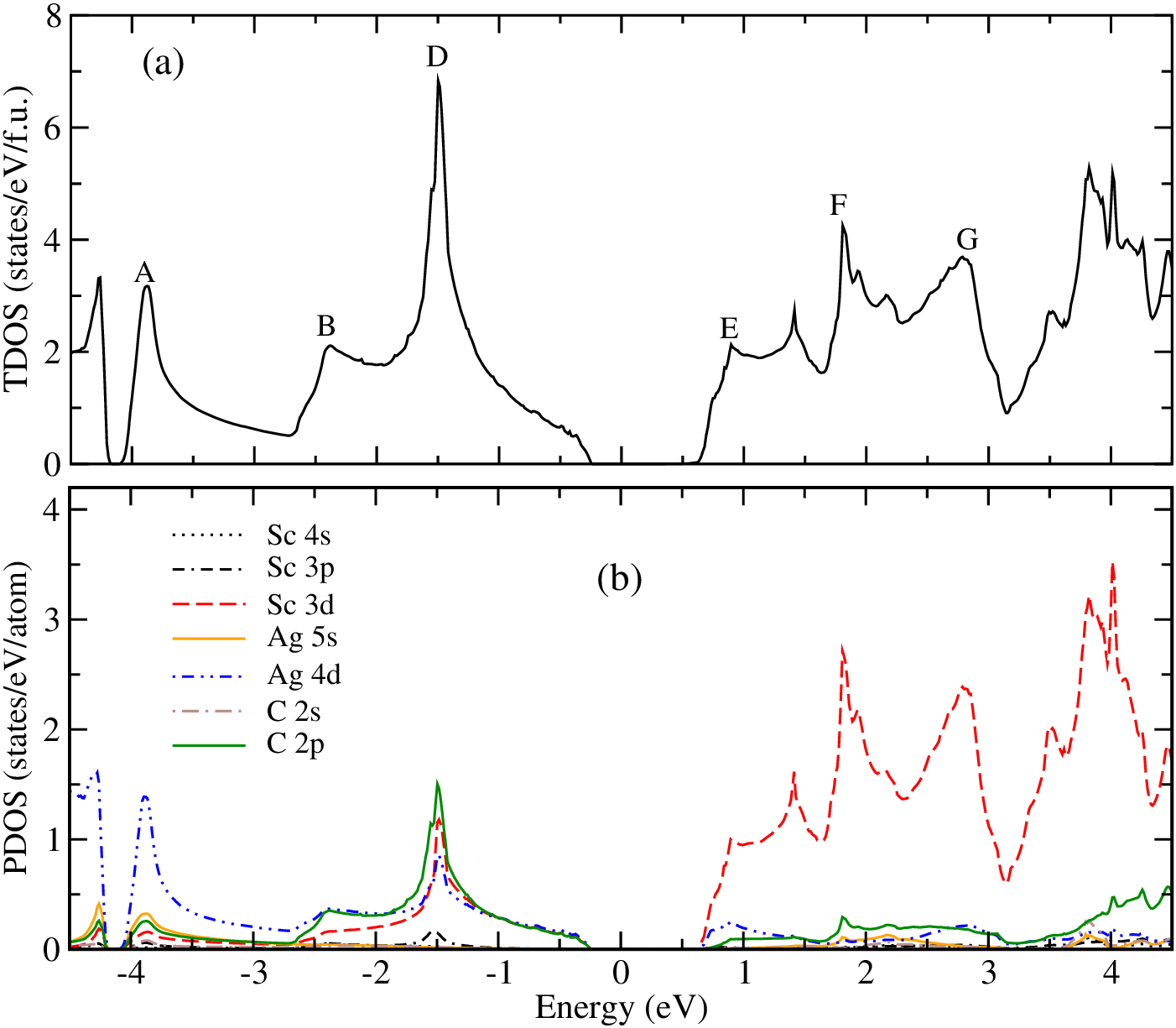} 
\caption{(a) Total and (b) partial density of states (T/PDOS) for the ScAgC compound.} 
\end{figure} 

\subsection{\label{sec:level2}Phonon properties}

Phonons have a significant role in deciding the thermal transport properties of materials. The phonon dispersion along the high symmetry axis in the first Brillouin zone for ScAgC is presented in Fig. 3. The obtained positive energies of phonons indicate that ScAgC is thermodynamically stable in the cubic phase with any small atomic displacements about the mean position. The phonon dispersion demonstrates the lattice vibration modes of ScAgC. All the phonons have a total of nine branches as the primitive cell contains three atoms. Among them, three modes have lower energy, which are acoustic phonons, and the rest, at higher energies, are optical phonons. Acoustic modes are mostly influenced by interactions between an atom in the unit cell and its counterpart atom in neighbouring unit cells, while interactions between two distinct atoms in the unit cell largely influence optical modes \cite{ashcroft1976}. The phonon dispersion curve can be separated into three zones based on this figure: the lower energy zone (LEZ), the middle energy zone (MEZ), and the upper energy zone (UEZ). The LEZ consists of three acoustic branches (one longitudinal acoustic (LA) and two transverse acoustic (TA)) and varies from 0 to $\sim$ 15.85 meV, in which two branches, S2 and S3 are degenerate along the $\Gamma$-$X$ and $\Gamma$-$L$ directions. Most of the heat transfer to a material happens through its acoustic modes of vibration. Similarly, there are six optical modes (two longitudinal optical (LO) and four transverse optical (TO)) in both the zones of MEZ and UEZ. Three optical modes are located in MFZ, ranging in energy from $\sim$ 24.5 meV to $\sim$ 34.2 meV, while the remaining three are located in UEZ and range from $\sim$ 44.35 meV to $\sim$ 54.20 meV. We have also seen from this figure that two optical modes are degenerate along the $X$-$\Gamma$-$L$ direction. At the $\Gamma$-point, the optical phonons in MEZ are triply degenerate at $\sim$ 24.5 meV. There exists a band-gap of $\sim$ 8.6 meV between LEZ (acoustic phonon) and MEZ (optical phonon). One can also notice that in the low energy region, the dispersion of acoustic phonons is almost linear, indicating that the phase velocity is the same as the group velocity.
\begin{figure}
\includegraphics[width=6cm, height=6cm]{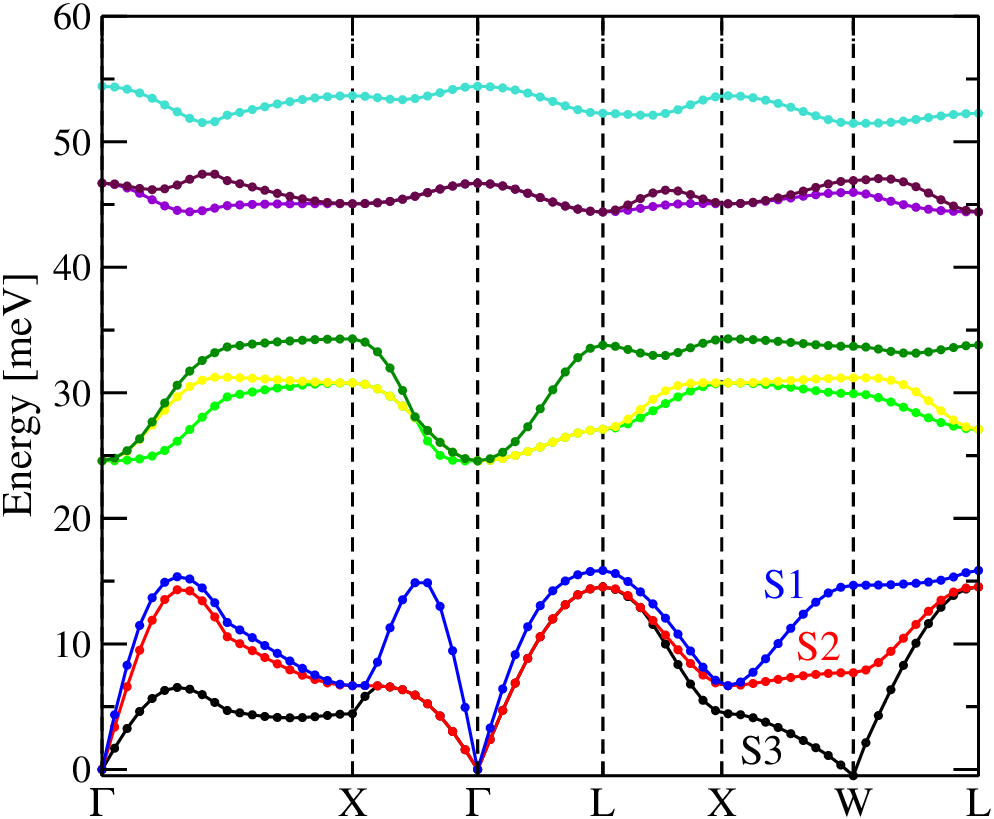} 
\caption{Phonon dispersion curve of the ScAgC compound.} 
\end{figure} 

The temperature beyond which all phonon modes start to be stimulated, and below which modes start to be frozen out, is known as the Debye temperature $\Theta_{D}$ \cite{ashcroft1976}. The maximum frequency of phonon is considered as Debye frequency $\omega_{D}$ in order to calculate $\Theta_{D}$. We can see from the phonon dispersion that the value of $\omega_{D}$ is $\sim$ 54.20 meV. Using the relation $\hbar\omega_{D}$ = $k_{B}$$\Theta_{D}$ \cite{ashcroft1976}, we have estimated $\Theta_{D}$ as $\sim$ 629 K. The obtained $\Theta_{D}$ of ScAgC is higher than the other HH compound like FeVSb ($\sim$ 547 K) \cite{shastri_FeVSb}. Further, the thermal properties of ScAgC have also been calculated. These properties play a big role in deciding the thermal response of a solid. The temperature-dependent lattice contributions to Helmholtz free energy $F$ and internal energy $E$ are calculated and illustrated in Figs. 4 (a) and (b). These figures show that $F$ and $E$ have non-zero positive values even when the temperature is zero, and this positive energy is the zero-point energy \cite{ashcroft1976}. This is because quantum particles exist in solids and vibrate, even at absolute zero temperature \cite{hao2012}. It can be understood through the asymptotic expressions of Eqs. (E1) and (E2) of supplementary material (SM) \cite{spple}:
\begin{eqnarray}
F_{0} = E_{0} = 3nN \int^{\omega_{L}}_0 \frac{\hbar\omega}{2} g(\omega) d\omega  
\end{eqnarray}  

The normalized phonon density of states $g$($\omega$) has been used, which means $\int^{\omega_{L}}_{0}$\textit{g}($\omega$)d$\omega$ = 1. 
\begin{figure}
\includegraphics[width=8cm, height=7cm]{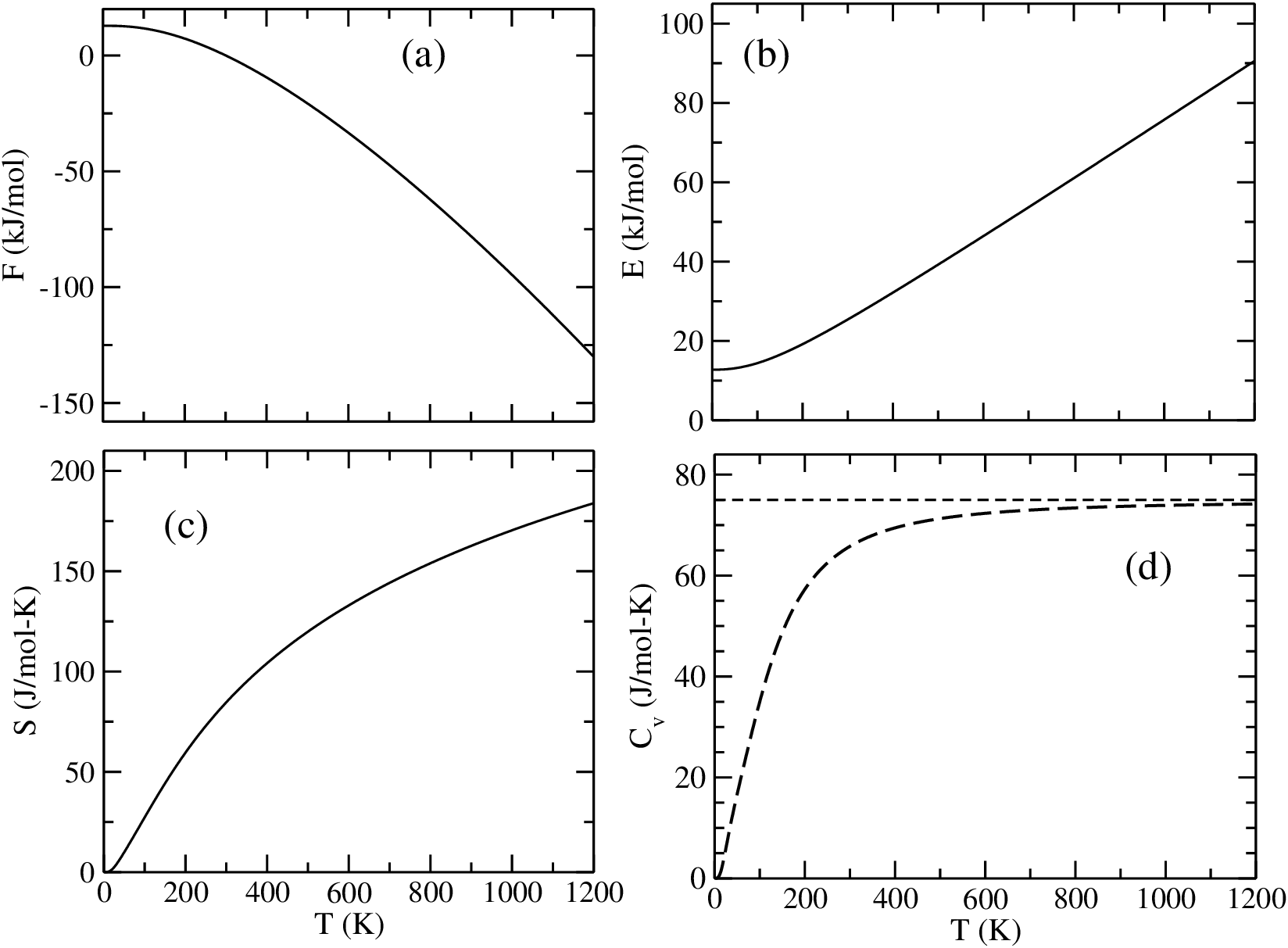} 
\caption{Lattice contributions to (a) the Helmholtz free energy $F$, (b) the internal energy $E$, (c) the entropy $S$, and (d) the constant-volume specific heat $C_{v}$ of the ScAgC compound.} 
\end{figure} 

From the plots, it can be seen that the zero-point energy of $F_{0}$ ($E_{0}$) is $\sim$ 12.26 (12.46) kJ/mol. After $\sim$ 200 K, there is a decrement in $F$ with increasing temperature. From $\sim$ 300 K to the highest studied temperature, $F$ exhibits negative values with a minimum value of $\sim$ -130 kJ/mol at 1200 K. In contrast, the change in their internal energy $E$ behaves in the opposite manner. From the Fig. 4(c), it is clear that entropy increases smoothly as temperature increases. The maximum calculated value of $S$ is $\sim$ 183.8 J$mol^{-1}$$K^{-1}$ at 1200 K. The phonon contribution to $C_v$ as a function of temperature is plotted in Fig. 4(d). We can notice that after $\Theta_{D}$ of $\sim$ 629 K, the $C_{v}$ curve approaches the classical Dulong and Petit limit \cite{ashcroft1976}, and that below $\Theta_{D}$, $C_{v}$ falls short of the Dulong and Petit limit, as the theoretical approach of $C_{v}$$\propto$ $T^{3}$ \cite{ashcroft1976}. Finally, it approaches zero at zero temperature.

\subsection{\label{sec:level2}Optical properties}

Aside from electronic properties, optical properties must also be considered when evaluating the performance of absorber materials \cite{fox2002}. The phenomena that arise from the intraction of light into semiconducting materials are covered by the optical properties of the materials. All the optical parameters have been assessed within DFT. But, we take a rigid shift in energy up to the $G_{0}W_{0}$ band-gap value of $\sim$ 1.01 eV in the absorption spectrum because this property is very sensitive to the material's band-gap. The DFT theory does not accurately describe the properties of materials in their excited states, such as band-gap and absorption spectrum \cite{yu2013}. All the optical parameters are estimated against the photon energy range of 0–10 eV.

As per the Kramers-Kronig relation, the real $\sigma_{1}(\omega)$ and imaginary $\sigma_{2}(\omega)$ parts of optical coductivity are important parameters for calculating the other optical properties of the material. It is known that electrons excite from VB to CB via absorbing photons, and this band-to-band (interband) transition is known as optical conduction \cite{ashcroft1976}. Fig. 5 presents the variation of optical conductivity $\sigma(\omega)$ with the photon energy. The value of $\sigma_{1}(0)$ is zero, indicating that the compound is a semiconductor \cite{Dressel2003}. In this figure, $\sigma_{1}(\omega)$ begins at a critical energy point of $\sim$ 0.3 eV. $\sigma_{1}(\omega)$ increases with an increase in photon energy up to $\sim$ 2.6 eV in the visible region with a corresponding value of $\sim$ $7.0\times 10^{3} (\Omega-cm)^{-1}$. ScAgC should be more conductive in the energy range of $\sim$ 1.7 eV to $\sim$ 2.6 eV. The increment in optical conductivity is due to the high absorption of the material. There are two prominent peaks having the values of $\sim$ 7$\times 10^{3}$ and $\sim$ 7.5$\times 10^{3}$ ($\Omega-cm)^{-1}$ at $\sim$ 2.6 and $\sim$ 5.3 eV, respectively. $\sigma_{2}(\omega)$ exhibits negative values up to $\sim$ 3.5 eV and both negative and positive values in the energy range of $\sim$ 3.5–5.3 eV. From $\sim$ 5.3 eV to highest studied energy, $\sigma_{2}(\omega)$ shows positive values with a maximum value of $\sim$ 7.8$\times 10^{3}$ ($\Omega-cm)^{-1}$ at $\sim$ 10 eV.
\begin{figure}
\includegraphics[width=5cm, height=6cm]{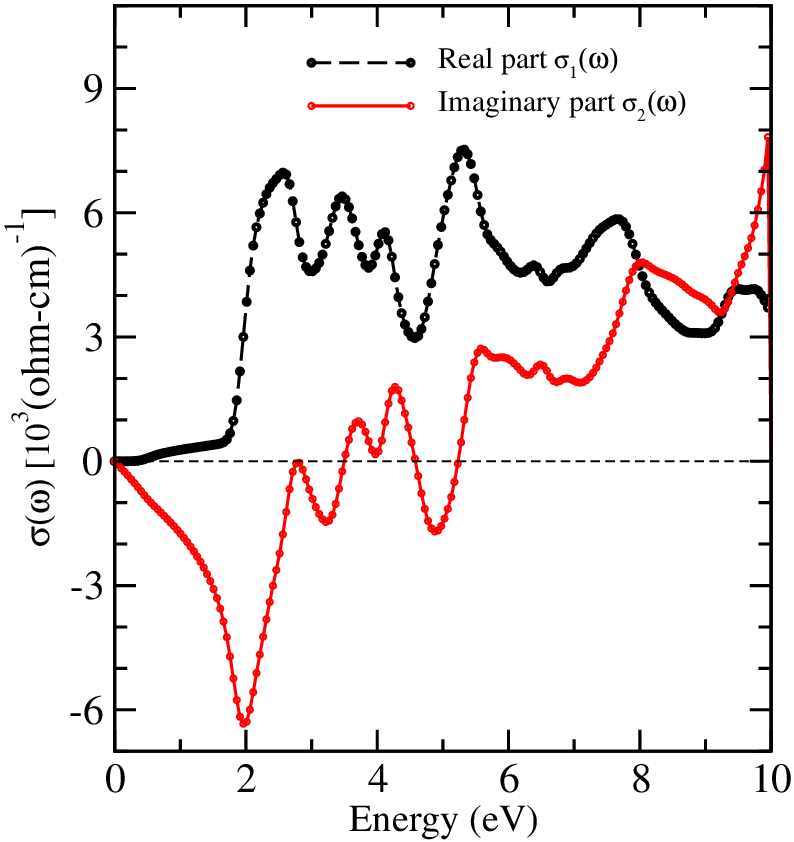} 
\caption{Real $\sigma_{1}(\omega)$ and imaginary $\sigma_{2}(\omega)$ part of optical conductivity $\sigma(\omega)$.} 
\end{figure}

The dielectric properties of a compound are generally decided by the frequency dependent dielectric function $\epsilon(\omega)$, which is mathematically given in Eqs. (E8) and (E9) in SM \cite{spple}. When materials interact with an electric field, molecules or crystals become polarized. Real part $\epsilon_{1}(\omega)$ describes their ability to store and remit energy, implying that no energy is lost during the polarization process \cite{Kasap2006}. But there will always be some losses in the medium during all polarization processes. The jumps of electrons between the two allowable states are critical for the interaction of light with the semiconductor, which gives us the optical absorption of the material. This information can be understood by imaginary part $\epsilon_{2}(\omega)$, which represents energy losses in the medium during the polarization process \cite{ambrosch2006,Kasap2006}. This concept is useful to understand the band characteristics, which means which band contributes more to the $\epsilon_{2}(\omega)$ spectrum's peaks. 

The $\epsilon_{1}(\omega)$ and $\epsilon_{2}(\omega)$ for ScAgC are presented in Fig. 6(a). The static dielectric constant $\epsilon_{1}$(0) is calculated to be $\sim$ 13.4, which is a reasonable agreement with another PV material like Si \cite{philipp1963}. The region from $\sim$ 1 eV to $\sim$ 1.95 eV is known as the normal dispersion region, in which $\epsilon_{1}(\omega)$ increases as energy increases \cite{fox2002}. $\epsilon_{1}(\omega)$ has a maximum peak value of $\sim$ 24.98 at $\sim$ 1.95 eV and immediately drops to $\sim$ 1.1 at $\sim$ 2.8 eV. The region of the energy range of $\sim$ 1.95–2.8 eV is called the anomalous dispersion region, in which $\epsilon_{1}(\omega)$ decreases as energy increases. In this region, the amplitude of the oscillation of electrons is relatively high, and thus a correspondingly huge amount of energy is dispersed \cite{fox2002}. $\epsilon_{1}(\omega)$ becomes zero at energy points of $\sim$ 3.57 eV, $\sim$ 3.87 eV, $\sim$ 4.07 eV, $\sim$ 4.5 eV and $\sim$ 5.3 eV. These energy points are called plasmon frequencies \cite{Dressel2003}. Also, $\epsilon_{1}(\omega)$ turns into some negative regions. The negative value shows that ScAgC acts like a metal and confirms that this compound reflects high energy photons \cite{ashcroft1976}. From the vector wave equation of $\omega^{2}$$\epsilon$ = $c^{2}$($K\cdot K$), the wave vector $K$ becomes imaginary number when $\epsilon_{1}$\textless 0 \cite{wang2019}. The $\epsilon_{1}(\omega)$ has lower values at higher energies (above $\sim$ 6 eV) of the studied energy range, indicating less interaction of this compound with the photon energy. The first main peak in the $\epsilon_{2}(\omega)$ spectra is located at $\sim$ 2.36 eV, the second one at $\sim$ 3.42 eV, while the third and fourth are respectively situated at $\sim$ 4.07 eV and $\sim$ 5.28 eV. These respective peaks come from band to band transitions of D$\rightarrow$E, D$\rightarrow$F, A$\rightarrow$E, and B$\rightarrow$G in Fig. 2(a). The results suggest that this compound has a high dielectric constant in the visible energy range and is well consistent with other solar materials like Si, GaAs, and InAs \cite{philipp1963}. 
\begin{figure}
\includegraphics[width=8cm, height=5cm]{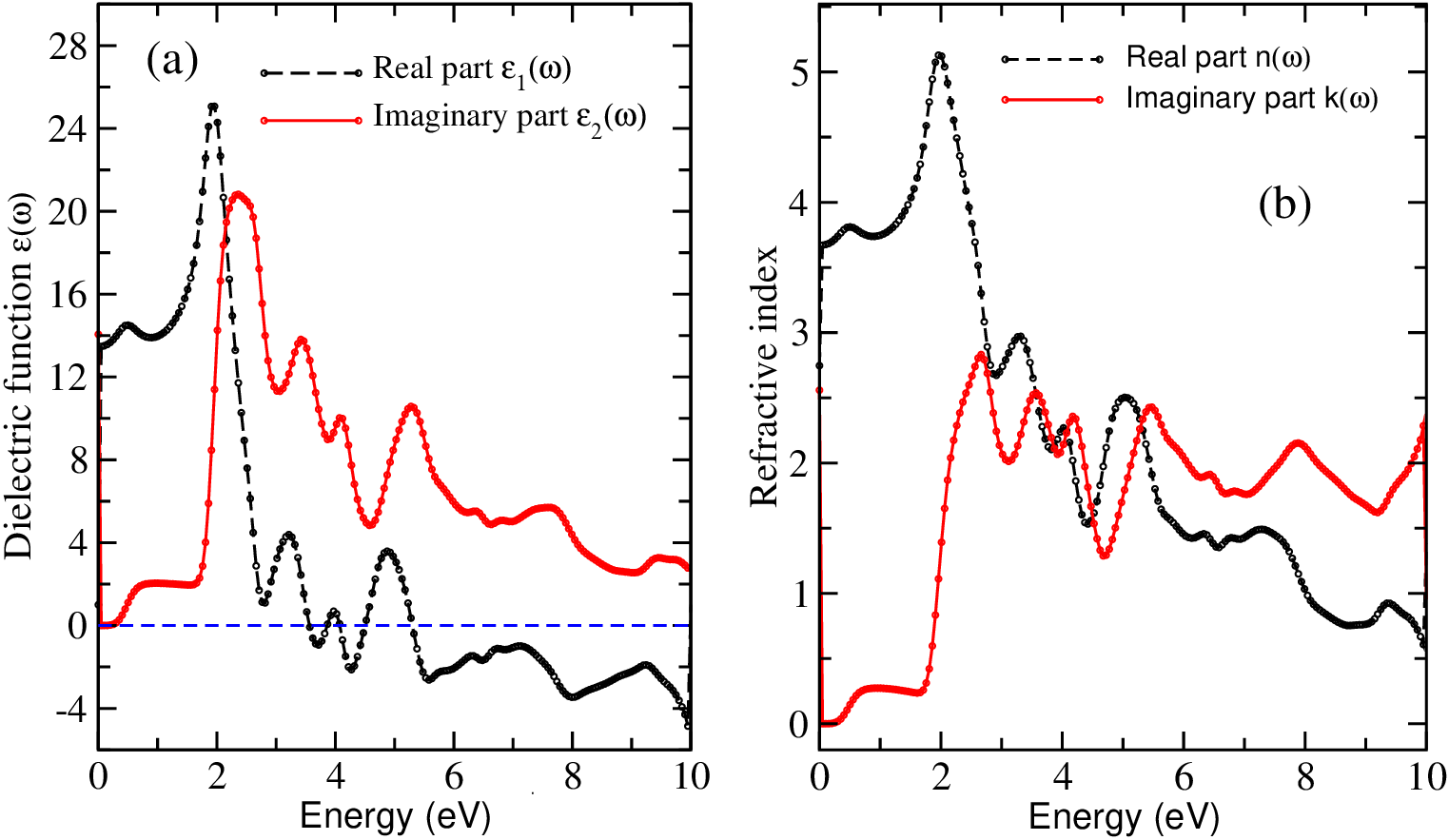} 
\caption{Complex (a) dielectric function $\epsilon(\omega)$ and (b) refractive index of ScAgC.} 
\end{figure}

The understanding of refractive index is essential for using materials in energy applications such as solar cells, LEDs, $etc$. The transparency of a compound to incident light is determined by its refractive index. The calculated real $n(\omega)$ and imaginary $k(\omega)$ parts of the refractive indices are shown in Fig. 6(b), which are given mathematically by Eqs. (E12) and (E13) of SM \cite{spple}. It has been observed that the $n(\omega)$ follows the same pattern as $\epsilon_{1}(\omega)$. The static refractive index $n(0)$ is estimated to be $\sim$ 3.7, while the $n(\omega)$ increases with photon energy in the IR region and decreases monotonically with a maximum value of $\sim$ 5.12 at $\sim$ 2 eV in the visible and UV regions and then it generally decreases. The minimum value of $n(\omega)$ is found to be $\sim$ 0.63 at $\sim$ 10 eV. A higher value of $n(\omega)$ represents the slower process of photons entering the material. This process happens due to interruptions in the path of travelling photons or interactions with electrons in a material. The spectra of the extinction coefficient $k(\omega)$ contains information about the light absorption measurements within an energy regime. The $k(\omega)$ spectra has several peaks and follows a similar pattern to the $\epsilon_{2}(\omega)$ spectra, as shown in Fig. 6(b). The peaks of the refractive index show nonlinear behaviour in the given photon energy range. $k(\omega)$ has a maximum intensity of $\sim$ 2.82 at $\sim$ 2.66 eV photon energy, which means photons will absorb very quickly at this energy. Both $n(\omega)$ and $k(\omega)$ start at maximum and minimum values, respectively, and meet at about 3 eV. Thereafter, they behave in an almost similar way up to 10 eV of photon energy. The value of $k(\omega)$ is greater than the value of $n(\omega)$ in the energy range from $\sim$ 5.3 eV to 10 eV, indicating that light cannot travel through this region. The studied optical properties $\sigma(\omega)$, $\epsilon(\omega)$, and $\tilde{n}(\omega)$ show a strong transition in the visible energy range, which means there can be strong electron-photon interaction. Due to this interaction, a greater number of electrons could be excited from the valence band to the conduction band. These properties suggest that ScAgC can be used in optical devices such as solar photovoltaics.

The $G_{0}W_{0}$ band-gap ($\sim$ 1.01 eV) of this compound lies in the ideal band-gap range ($\sim$ 1-1.5 eV) for solar cell performance \cite{shockley1961}. However, having a perfect band-gap doesn't always ensure that the material will efficiently absorb light. Thus, the absorption coefficient $\alpha(\omega)$ provides valuable information about the material's light absorption, which is an essential aspect in calculating the efficiency of solar materials. Basically, $\alpha(\omega)$ indicates whether a material is metallic, an insulator, or a semiconductor in terms of its electronic nature. The plot of $\alpha(\omega)$ in Fig. 7(a) shows that there is no absorption up to the $G_{0}W_{0}$ band-gap of $\sim$ 1.01 eV. Absorption starts after $\sim$ 1.01 eV and slowly increases up to $\sim$ 2.3 eV. After $\sim$ 2.3 eV, our result shows a rapid increment of $\alpha(\omega)$ with increasing photon light up to $\sim$ 3.25 eV. A sharp absorption peak of $\sim$ 0.7 $\times 10^{6}$ $cm^{-1}$ has been found at $\sim$ 3.25 eV in the visible region. The figure shows the variation of $\alpha(\omega)$ in this studied energy range, which means the material behaves as an opaque compound. We have found the first peak of $\sim$ 1.34 $\times 10^{6}$ $cm^{-1}$ at $\sim$ 6 eV in the UV region. The highest estimated value of $\alpha(\omega)$ is $\sim$ 1.7 $\times 10^{6}$ $cm^{-1}$ at $\sim$ 8.5 eV of photon energy. Furthermore, reflectivity also gives better information to decide whether a material has excellent potential in the solar field or not. From Eq. (E15) of SM \cite{spple}, we need both $n(\omega)$ and $k(\omega)$ to calculate the $r(\omega)$ of the material. The resulting plot, depicted in Fig. 7(b), has a reflectivity of $\sim$ 0.33 at $\sim$ 1 eV of visible energy. Also, at $\sim$ 4.7 eV, $r(\omega)$ has a minimum value of $\sim$ 0.24. Hence, the major part of the light is absorbed in the vicinity of $\sim$ 4.7 eV. $r(\omega)$ spectra behave similarly to $\sigma_{1}(\omega)$ spectra with photon energy variations. Moreover, it is also seen that the maximum peak of $r(\omega)$ occurs in the energy beyond $\sim$ 8 eV, in which the $\epsilon_{1}(\omega)$ lies in a negative region, which tells us the reflection of the light from the material surface.
\begin{figure}
\includegraphics[width=8cm, height=5cm]{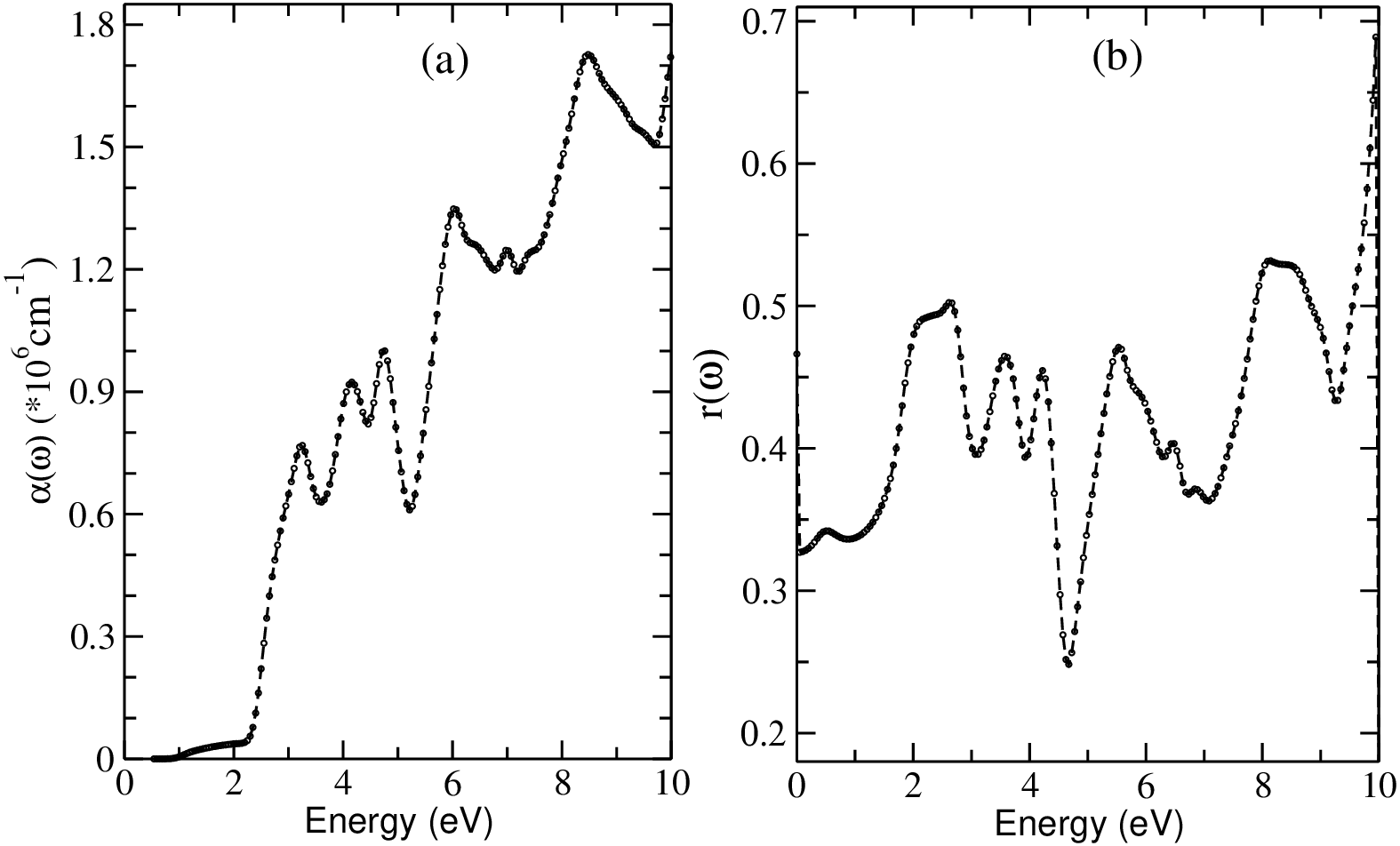} 
\caption{ Calculated (a) absorption coefficient $\alpha(\omega)$ and (b) reflectivity $r(\omega)$ of the ScAgC compound.}  
\end{figure}
\begin{figure}
\includegraphics[width=5.5cm, height=5cm]{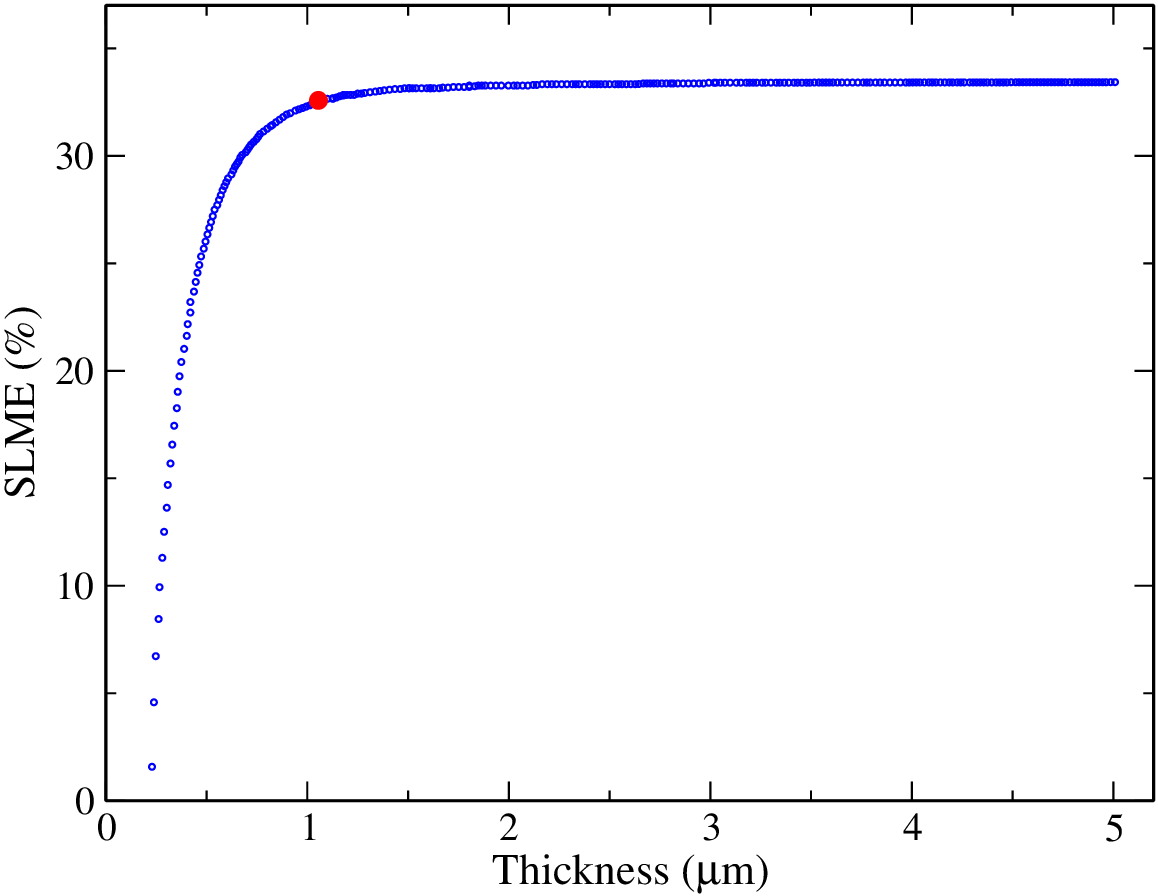} 
\caption{ The SLME with respect to thin film thickness ($L$) for ScAgC.}  
\end{figure}

Appropriate estimation of the efficiency of PV materials is necessary for material research and innovation. HH compounds have great potential in the development of solar cell technology with rapid increases in their efficiency. This is due to its high absorption capacity and long carrier diffusion length. The SLME depends upon the material's band-gap due to the Shockley and Queisser (SQ) limit \cite{shockley1961}, which selects good materials with their band-gap (direct or indirect) for PV applications. However, strong optical absorption above a suitable band-gap is essential to achieve high efficiency for solar materials \cite{yu2013}. The $G_{0}W_{0}$ band-gap and the shape of the absorption spectra are used to calculate the SLME at 300 K. According to the SLME definition in SM \cite{spple}, it is also affected by the thin film thickness ($L$) of material. The effect of the ScAgC thickness on the efficiency is clearly shown in Fig. 8. Due to weak absorption, SLME is small at very low $L$. After this, SLME goes up as $L$ goes up, with a maximum value of about 33\% at $\sim$ 1 $\mu m$. At very large $L$, SLME approaches the SQ limit \cite{ruhle2016} and meets the requirement that the material with a band-gap closer to 1.1 eV has a large SLME \cite{choudhary2019}. The SLME of this compound is bigger than that of other HH compounds that have been studied in the earlier literatures \cite{sahni2020,zhang2012,choudhary2019}. If we include the material-property based spectroscopic parameters in SLME, we should choose an appropriate size of $L$. Therefore, $\sim$ 1 $\mu$m thickness of ScAgC can be used as a solar material to get an optimum SLME.       

\subsection{\label{sec:level2}Thermoelectric properties}
As from electronic properties, the sharp DOS at the band edge and the width of the band-gap are critical characteristics that influence the TE properties of the ScAgC compound. The presence of a semiconducting ground state with an almost flat CB confirmed that this compound would have good TE properties. The efficiency of TE material is decided by its $PF$. Raising the $PF$ is one of the ways to increase the efficiency of TE materials. The $G_{0}W_{0}$ band-gap is used to calculate the $PF$ of this considered material as a function of $\mu$ at different temperatures. The $PF$ with respect to $\mu$ ranging from -1800 meV to 1800 meV at different temperatures of 300 K to 1200 K is shown in Fig. 9. In this plot, the positive and negative values of $\mu$ represent the electron and hole doping in the material, respectively. It can be seen that the $PF$ greatly increases with the increase in temperature. Also, this plot contains two peaks, corresponding to each temperature in the considered range of $\mu$. The maximum values of $PF$ at $\mu$ of $\sim$ 1000 meV are calculated as $\sim$ 145$\times 10^{14}$ $\mu WK^{-2}cm^{-1}s^{-1}$ and $\sim$ 19$\times 10^{14}$ $\mu WK^{-2}cm^{-1}s^{-1}$ at 1200 K and 300 K, respectively. These $PFs$ correspond to electron concentrations of $\sim$ 3.9$\times$$10^{21}$$cm^{-3}$ and $\sim$ 0.9$\times$$10^{21}$$cm^{-3}$, respectively. Similarly, at $\sim$ 660 meV, the maximum values of $PF$ are estimated to be $\sim$ 123 $\times 10^{14}$ $\mu WK^{-2}cm^{-1}s^{-1}$ and $\sim$ 15 $\times 10^{14}$ $\mu WK^{-2}cm^{-1}s^{-1}$ at 1200 K and 300 K, respectively. The optimal hole doping concentration containing maximum $PF$ is $\sim$ 1.5$\times$$10^{21}$$cm^{-3}$ and $\sim$ 2.04$\times$$10^{21}$$cm^{-3}$, respectively. We can see from the figure that $PF$ for $n$-type doping is higher than $p$-type doping in this material. This signifies that the $n$-type could have superior thermoelectric performance than the $p$-type. It can also be noticed that as the temperature increases from 300 K to 1200 K, all peaks shift slightly towards the lower (higher) magnitude of $\mu$ corresponding to negative (positive) $\mu$. The high value of $PF$ for this compound can also be understood by their DOS. As we have observed from Fig. 2, this compound has a high DOS value in the vicinity of the Fermi level ($E_{F}$).
\begin{figure}
\includegraphics[width=6.5cm, height=5.5cm]{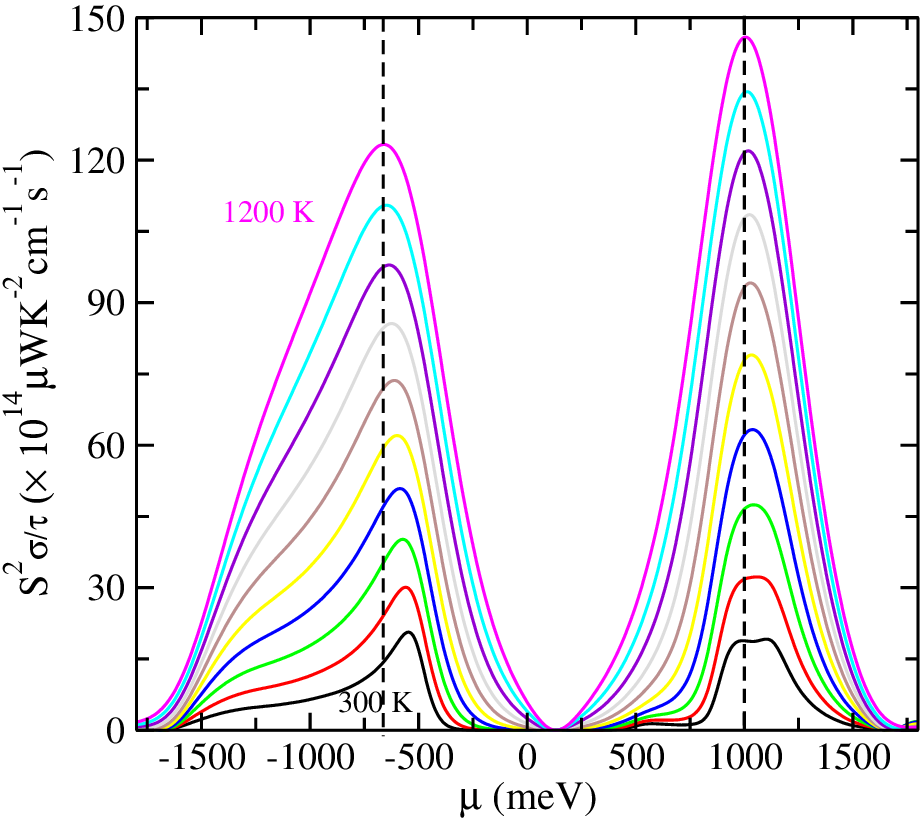} 
\caption{ Power factor ($PF$) per relaxation time with the variation of chemical potential ($\mu$) at different temperatures.}  
\end{figure}

The calculation of the total thermal conductivity $\kappa$ of a TE material is an important part of fully evaluating the TE materials. But, the calculation of the lattice part of $\kappa$ ($\kappa_{ph}$) necessitates a large computational cost \cite{togo2015}. So, any simple method that can predict a reasonably accurate value of $\kappa_{ph}$ is much more helpful. Here, we calculate phonon relaxation time ($\tau_{ph}$) using the simple method. For this, we consider two Heusler compounds that have experimentally different $\kappa_{ph}$ value. Then, $\tau_{ph}$ of first Heusler compound has been used to calculate $\kappa_{ph}$ of second compound and compared it with the experimental value, which was found to be reasonably good matched. Furthermore, we used this $\tau_{ph}$ in our calculation. The details of this simple method can be found in the work of Shastri \textit{et al.} \cite{shastri2019}.
 
\begin{eqnarray}
\kappa_{ph}= \frac{1}{3}C_{v}c^{2}\tau_{ph}
\end{eqnarray}  
Where $C_{v}$ is the specific heat per unit volume at constant-volume, $c$ is the phase velocity of phonon in linear dispersion.

\begin{table}
\caption{\small{Phase velocity ($c$) of three acoustic phonons (S1–S3) for ScAgC compound in $\Gamma$–$L$ and $\Gamma$–$X$ directions and average of phase velocities $c_{avg}$ in $THz\cdot Bohr$}}
\resizebox{0.34\textwidth}{!}{%
\setlength{\tabcolsep}{6pt}
\begin{tabular}{@{\extracolsep{\fill}}c c c c c  } 
\hline\hline
 
                               
   & & \multicolumn{1}{c}{$\Gamma$-$L$} &&   \multicolumn{1}{c}{$\Gamma$-$X$}     \\
     
 \hline
S1 &&   48.37  && 63.97    \\
S2 &&   38.86  && 24.14   \\
S3 &&   38.86  && 24.14   \\
\hline
$c_{avg}$ &  & &39.73  \\
\hline\hline
\end{tabular}}
\end{table}

We can obtain $C_{v}$ as a function of temperature using Fig. 4(d). In order to calculate $c$, the linear dispersion of three acoustic modes near the $\Gamma$-point is used to fit a straight line in the calculated phonon dispersion shown in Fig. 3. Thus, $c$ is obtained as $c$ = $\frac{\omega}{k}$, which is the slope of the dispersion curve. The acoustic modes in the $\Gamma$-$L$ and $\Gamma$-$X$ directions were used to calculate $c$. In Eq. (2), instead of $c$, the average of the $c$ values, which we call $c_{avg}$, of three acoustic branches in two directions is used to calculate $\kappa_{ph}$. Table II shows the obtained values of $c$ and $c_{avg}$ for three acoustic branches in two directions for ScAgC. The branch notations S1, S2, and S3 are shown in the phonon dispersion curve. Table II shows that the $c$ values decrease in both directions from S1 to S2 branch, but remain same for S2 and S3. The value of $c$ for the S1 branch in the $\Gamma$-$L$ ($\Gamma$-$X$) direction is 48.37 (63.97) THz$\times$Bohr. Similarly, S2 and S3 are degenerate along the $\Gamma$-$L$ and $\Gamma$-$X$ directions, with $c$ values of 38.86 and 24.14 THz$\times$Bohr, respectively. The $c_{avg}$ in both directions for three acoustic branches is 39.73 THz$\times$Bohr.

Once we have $c$ and $C_{v}$, $\tau_{ph}$ needs to be computed in order to calculate $\kappa_{ph}$. The $\tau_{ph}$ is calculated as $\sim$ 3.1$\times$$10^{-15}$$s$ for FeVSb HH compound \cite{shastri_FeVSb} at 300 K using the simple method as discussed earlier. Further, we have obtained the value of $\kappa_{ph}$ from 300 K to 1200 K using this $\tau_{ph}$. From the Fig. 10, the $\kappa_{ph}$ value increases as the temperature increases. The value of $\kappa_{ph}$ at 300 K is $\sim$ 3.36 W$m^{-1}K^{-1}$, which is a low value when compared to other HH compounds \cite{shastri_FeVSb,shastri_ZrNiSn}. Initially, $\kappa_{ph}$ increases rapidly up to $\sim$ 600 K, then gradually increases, reaching a maximum value of $\sim$ 3.8 W$m^{-1}K^{-1}$ at 1200 K. It is important to note that as temperature increases, the relaxation time decreases, so one can expect the low value of $\kappa_{ph}$ at higher temperatures. Therefore, it is expected to have a high $ZT$ at high temperature region.

We determined $ZT$ at different temperatures for doped ScAgC with optimal hole and electron doping to evaluate the material for TE applications. A high value of $ZT$ can be obtained by a large $PF$ in the numerator and at the same time a low value of $\kappa$ in the denominator, as can be seen in Eq. (E28) of SM \cite{spple}. For this semiconducting material, we used the constant electron relaxation time $\tau$ values of 1$\times$$10^{-14}$$s$ ($\tau_{1}$) and 1$\times$$10^{-15}$$s$ ($\tau_{2}$) for the estimation of $\sigma$ and $\kappa_{e}$ \cite{ashcroft1976}. The calculated temperature-dependent $ZT$ values for $\tau_{1}$ and $\tau_{2}$ are shown in Fig. 11(a) for $n$-type and $p$-type doping of this compound. The line style in Fig. 11(b) indicates the different $\tau$ values used to calculate the $ZT$ of doped ScAgC. From the figure, one can notice that the $ZT$ values for hole ($p$-type) doping are greater than those for electron ($n$-type) doping for $\tau_{1}$, and they behave in the opposite manner for $\tau_{2}$. This is because the value of $\kappa_{e}$ for $\tau_{1}$ is comparatively higher for the $n$-type of ScAgC than for the $p$-type. At 1200 K, the maximum values of $ZT$ for $p$-type are $\sim$ 0.53 and $\sim$ 0.24 for $\tau_{1}$ and $\tau_{2}$, respectively. Similarly, the maximum $ZT$ for the $n$-type is $\sim$ 0.47 and $\sim$ 0.25 at 1200 K for the two successive $\tau$ values used, respectively.
\begin{figure}
\includegraphics[width=6cm, height=5.5cm]{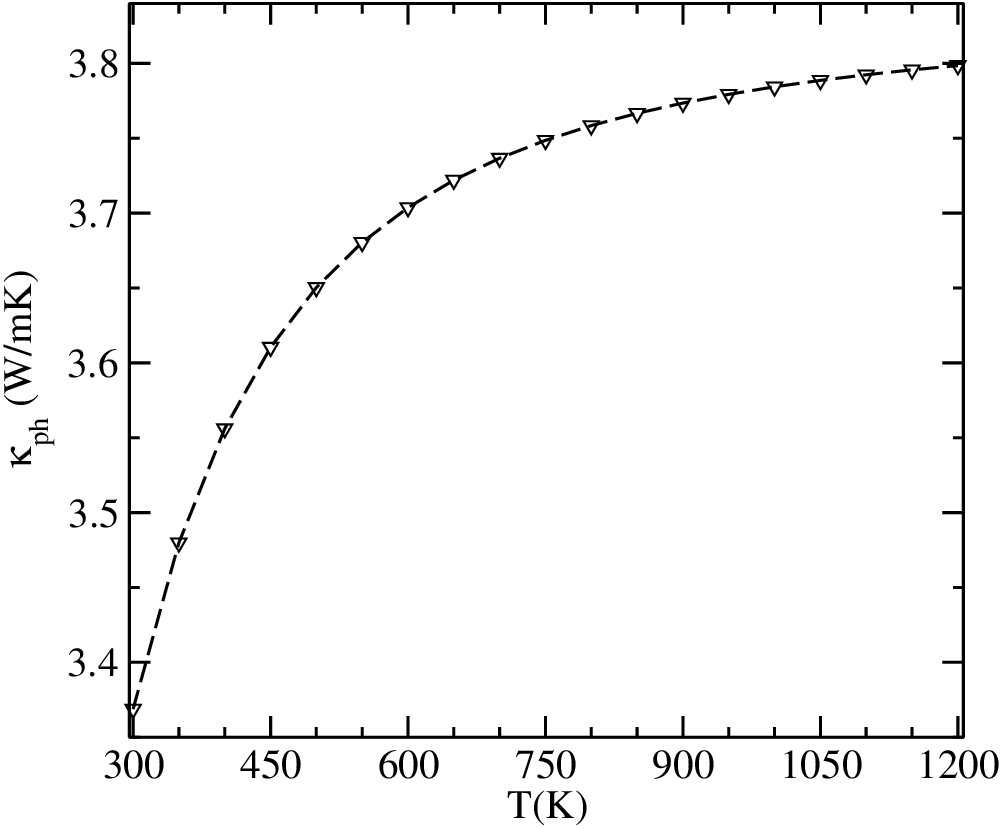} 
\caption{The calculated lattice part of thermal conductivity $\kappa_{ph}$ of ScAgC with temperature.}  
\end{figure}
\begin{figure}
\includegraphics[width=8.5cm, height=5cm]{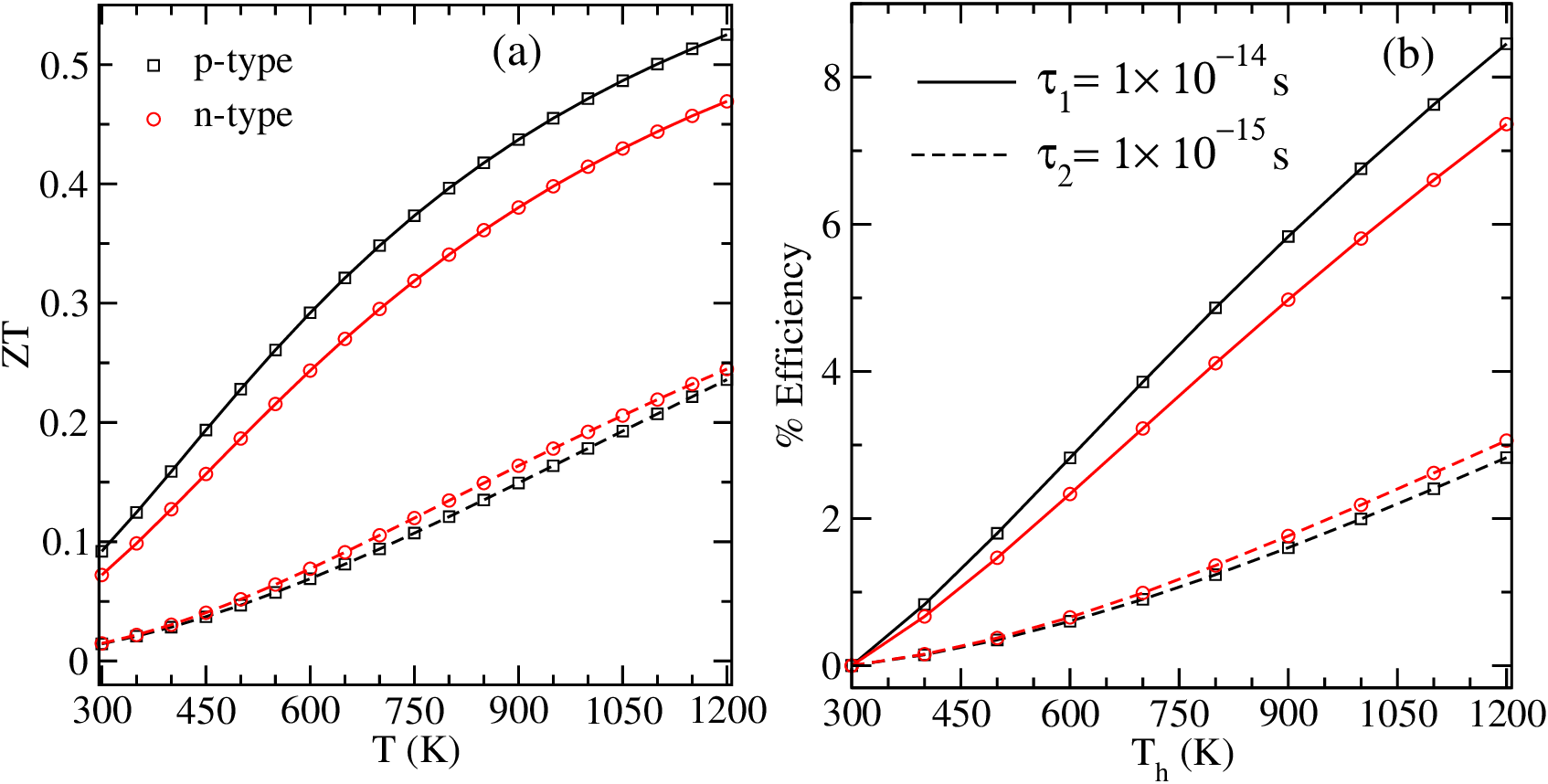} 
\caption{ (a) The calculated figure of merit $ZT$, and (b) the variation of the \% efficiency ($\eta$) with respect to hot temperature, keeping fixed cold side temperature at 300 K.}  
\end{figure}

It is preferable to calculate the conversion \% efficiency ($\eta$) of TE material once the $ZT$ is known. Here, the segmentation method has been considered for estimating the $\eta$ \cite{gaurav2017,sk2020,shastri_FeVSb}. Fig. 11(b) shows the calculated maximum possible $\eta$ for $n$-type and $p$-type compounds using $\tau_{1}$ and $\tau_{2}$. In order to estimate $\eta$, the cold side temperature ($T_{c}$) was kept at a constant temperature, while the hot side temperature ($T_{h}$) was varied from 300 K to 1200 K. From the figure, the values of $\eta$ calculated by $\tau_{1}$ ($\tau_{2}$) for $p$-type ($n$-type) compounds are greater than those calculated for $n$-type ($p$-type) compounds. For both types of doping, we can see that the value of $\eta$ increases as $T_{h}$ increases. When $T_{c}$ ($T_{h}$) is set at 300 K (1200 K), the optimum values of $\eta$ for $p$-type ScAgC are $\sim$ 8.5\% and $\sim$ 2.83\% for two consecutive $\tau$ values, respectively. Similarly, $n$-type ScAgC has a maximum $\eta$ of $\sim$ 7.4\% ($\tau_{1}$) and $\sim$ 3.1\% ($\tau_{2}$), when $T_{c}$ and $T_{h}$ are set to be at 300 K and 1200 K, respectively. The conversion efficiency of ScAgC suggests that this compound is competitive with those of other HH compounds such as doped TaFeSb \cite{zhu2019}, FeNbSb \cite{fu2015}, and ZrCoBi \cite{zhu2018} for TE applications. 

This study demonstrates that the high values of $\sigma (\omega)$, $\epsilon (\omega)$, $\tilde{n}(\omega)$, $\alpha (\omega)$, and SLME, as well as the low r($\omega$) of this proposed compound, fall into the category of all the necessary conditions for PV applications. In the high temperature region, Si-Ge based alloys having $ZT$ ranges of 0.4-0.6 for $p$-type and 0.5-0.8 for $n$-type are one of the good candidates for TE applications \cite{snyder2008}. In this competition, doped ScAgC could be a promising candidate for high temperature power generation. We believed that this study could provide a useful theoretical understanding for the experimental characterization of this compound.
 
\section{Conclusions}
In summary, the electronic properties of ScAgC as well as detailed analysis of the optical and TE properties have been investigated using first-principles calculations and semi-classical Boltzmann transport theory. This compound shows a direct band-gap of $\sim$ 0.47 eV calculated from DFT, while the band-gap using $G_{0}W_{0}$ method is $\sim$ 1.01 eV. The calculated phonon dispersion has no soft modes, which means ScAgC is a stable compound, and phonon-contributed thermal properties are also investigated. The optical parameters such as $\sigma(\omega)$, $\epsilon(\omega)$, and $\tilde{n}(\omega)$ are more likely to absorb light in the visible region. Furthermore, we observed a strong absorption of $\sim 1.7\times$$10^{6}$ $cm^{-1}$ at $\sim$ 8.5 eV and a low reflectivity of $\sim$ 24\% at $\sim$ 4.7 eV for this compound. To check the performance of the material for PV applications, the SLME is calculated at 300 K and shows a maximum value of $\sim$ 33\% at $\sim$ 1 $\mu m$ thickness. The $\kappa_{ph}$ shows maximum (minimum) value of $\sim$ (3.36) 3.8 W$m^{-1}K^{-1}$ at 300 (1200) K. At 1200 K, the electron-doped ScAgC shows a maximum $PF$ of $\sim$ 145 $\times$ $10^{14}$ $\mu WK^{-2}cm^{-1}s^{-1}$ for $\sim$ 3.9 $\times$ $10^{21}$$cm^{-3}$ of the concentration. At the same temperature, for hole-doped ScAgC with a concentration of $\sim$ 1.5 $\times$ $10^{21}$$cm^{-3}$, the optimum $PF$ is predicted to be $\sim$ 123 $\times$ $10^{14}$ $\mu WK^{-2}cm^{-1}s^{-1}$. The maximum $ZT$ at 1200 K is estimated to be 0.53, while the optimum \%efficiency is 8.5\% for cold and hot temperatures of 300 K and 1200 K, respectively. This research suggests that ScAgC can be used in solar cell absorbers and TE applications for clean energy technology.

\section{References}

\bibliography{MS}
\bibliographystyle{apsrev4-2}

\pagebreak
\begin{center}
\textbf{Supplementary material for \textquotedblleft First-principles study of optoelectronic and thermoelectronic properties of the ScAgC half-Heusler compound\textquotedblright}
\end{center}

\subsection{\label{sec:level2}Phonon properties}
\setcounter{equation}{0}
\setcounter{figure}{0}
\renewcommand{\theequation}{E\arabic{equation}}
\renewcommand{\thefigure}{S\arabic{figure}}
Within the harmonic approximation \cite{maradudin1963}, the thermodynamical properties such as the lattice contribution to the Helmholtz free energy $F$, the phonon contribution to the internal energy $E$, as well as the constant-volume specific heat $C_{v}$ and the entropy $S$ at temperature $T$ are expressed as follows \cite{lee1995}:
\begin{eqnarray}
F = 3nNk_{B}T \int^{\omega_{L}}_{0} \ln \{2\sinh\frac{\hbar\omega}{2k_{B}T}\} g(\omega)d\omega,  
\end{eqnarray}

\begin{equation}
E = 3nN\frac{\hbar}{2} \int^{\omega_{L}}_{0} \omega \coth (\frac{\hbar\omega}{2k_{B}T}) g(\omega)d\omega,  
\end{equation}

\begin{eqnarray}
C_{v} = 3nNk_{B}\int^{\omega_{L}}_{0}(\frac{\hbar\omega}{2k_{B}T})^{2}\csc h^{2}(\frac{\hbar\omega}{2k_{B}T}) \nonumber \\
\times g(\omega)d\omega,   
\end{eqnarray}

\begin{eqnarray}
S = 3nNk_{B}\int^{\omega_{L}}_{0}[\frac{\hbar\omega}{2k_{B}T}\coth(\frac{\hbar\omega}{2k_{B}T}) - \nonumber \\
\ln\{2\sinh\frac{\hbar\omega}{2k_{B}T}\}]g(\omega)d\omega   
\end{eqnarray}
Where $k_{B}$, $n$, $N$, $g$($\omega$) and $\omega_{L}$ are the Boltzmann constant, the number of atoms per unit cell, the number of unit cells, phonon density of states, and the highest phonon frequency, respectively.

\subsection{\label{sec:level2}optical properties}
The frequency-dependent conductivity $\sigma(\omega)$ does have real and imaginary parts.
\begin{eqnarray}
\sigma(\omega) = \sigma_{1}(\omega)+\iota\sigma_{2}(\omega)
\end{eqnarray}
The real part of the electrical conductivity $\sigma_{1}(\textbf{k},\omega)$ can be obtained using the Kubo–Greenwood formulation as \cite{callaway2013,mazevet2010},
\begin{eqnarray}
\sigma_{1}(\textbf{k},\omega)=\frac{2\pi}{3\omega\Omega}\sum^{n_{b}}_{j=1}\sum^{n_{b}}_{i=1}\sum^{3}_{\alpha=1}(F(\epsilon_{i,\textbf{k}})-F(\epsilon_{j,\textbf{k}}))\times \nonumber \\
\mid \textless \Psi_{j,\textbf{k}}\mid \nabla_{\alpha}\mid\Psi_{i,\textbf{k}}>\mid ^{2} \delta(\epsilon_{j,\textbf{k}}-\epsilon_{i,\textbf{k}}-\omega)  
\end{eqnarray}
Here, $\omega$ is the frequency of light. The Planck's constant $\hbar$, electron charge $e$, and the mass of electron $m_{e}$ are all set to one in atomic units. All bands up to $n_{b}$ (by $i$ and $j$) and three spatial directions (by $\alpha$) are involved in the triply periodic calculation of the cubic supercell volume element $\Omega$. The weight of the specific \textbf{k}-point related to the $i^{th}$ band energy of $\epsilon_{i,\textbf{k}}$ is denoted by $F(\epsilon_{i,\textbf{k}})$. The Kohn-Sham wave function of the $i^{th}$ band with a perticular \textbf{k}-point is $\Psi_{i,\textbf{k}}$. The conservation of total energy is conferred by the term of $\delta(\epsilon_{j,\textbf{k}}-\epsilon_{i,\textbf{k}}-\omega)$. We can obtain total optical conductivity by summing over all the necessary \textbf{k}-points. 

It is also possible to calculate the frequency-dependent imaginary part of optical conductivity $\sigma_{2}(\omega)$ using $\sigma_{1}(\omega)$ from the Kramers-Kronig \cite{callaway2013} relation as,
\begin{eqnarray}
\sigma_{2}(\omega) = \frac{-2P}{\pi} \int\frac{\sigma_{1}(\nu)\omega}{(\nu^{2}-\omega^{2})}d\nu 
\end{eqnarray}  
In the above expression, $P$ indicates the Cauchy principle value of the contour integral. Also, $\nu$ is related to the contribution of frequency along the real axis in the contour. The frequency-dependent real $\epsilon_{1}(\omega$) and imaginary $\epsilon_{2}(\omega$) parts of the dielectric function $\epsilon(\omega)$ can be written using both parts of the $\sigma(\omega)$, which is given below,
\begin{eqnarray}
\epsilon_{1}(\omega) = 1 - \frac{4\pi}{\omega}\sigma_{2}(\omega),
\end{eqnarray}
\begin{eqnarray}
\epsilon_{2}(\omega) = \frac{4\pi}{\omega}\sigma_{1}(\omega),
\end{eqnarray}
The total dielectric function $\epsilon(\omega)$ is defined as,
\begin{eqnarray}
\epsilon(\omega) = \epsilon_{1}(\omega) + \iota\epsilon_{2}(\omega) = [n(\omega) + \iota k(\omega)]^{2}
\end{eqnarray}
 Where,
\begin{eqnarray}
\tilde{n}(\omega)= n(\omega) + \iota k(\omega)
\end{eqnarray}
$\tilde{n}(\omega)$ is known as the complex refractive index.

Finally, the other optical parameters such as the real $n(\omega)$ and imaginary $k(\omega)$ parts of refractive index, reflectivity $r(\omega)$, and absorption coefficient $\alpha(\omega)$ can be computed using the dielectric function, which are defined as,
\begin{eqnarray}
n(\omega) = \sqrt{\frac{1}{2}[\mid\epsilon(\omega)\mid + \epsilon_{1}(\omega)]},
\end{eqnarray}
\begin{eqnarray}
k(\omega) = \sqrt{\frac{1}{2}[\mid\epsilon(\omega)\mid - \epsilon_{1}(\omega)]},
\end{eqnarray}

\begin{eqnarray}
\alpha(\omega) = \frac{4\pi}{n(\omega)\times c}\sigma_{1}(\omega),
\end{eqnarray}
and,

\begin{eqnarray}
r(\omega) = \frac{[1 - n(\omega)]^{2} + k(\omega)^{2}}{[1 + n(\omega)]^{2} + k(\omega)^{2}}
\end{eqnarray}
Where $c$ is the speed of light.

Solar cells convert sunlight into electricity in the form of voltage and current via the photovoltaic (PV) process. The power conversion efficiency of a PV solar cell, SLME ($\eta$), is defined as the ratio of maximum output power density ($P_{max}$) to total incident solar energy density ($P_{in}$) \cite{choudhary2019}. $P_{max}$ can be calculated as the product of maximum voltage $V$ and maximum current density $J$. 
\begin{eqnarray}
\eta = \frac{P_{max}}{P_{in}} 
\end{eqnarray}

If the solar cell operates like an ideal diode at temperature $T$ and is turned on by photon flux $I_{sun}$, then the total current density $J$ is given by, 
\begin{eqnarray}
J = J_{sc} - J_{0}(e^{eV/kT} - 1)
\end{eqnarray}
Where $e$ denotes the elementary charge, $V$ is the voltage across the absorber layer, $J_{0}$ is the constant, and $k$ is the Boltzmann’s constant. The first term is the short-circuit current density $J_{sc}$, which is written as,
\begin{eqnarray}
J_{sc} = e\int^{\infty}_{0}a(E) I_{sun}(E) dE
\end{eqnarray}
Here $a(E)$ denotes the photon absorptivity, and for sunlight, $I_{sun}$ is the spectrum of the Air Mass (AM) 1.5G radiation \cite{AM1.5}. In this AM1.5G spectrum, solar cells are typically tested in laboratories. For the calculation of $a(E)$, we have taken a slab of compound of thickness $L$ with zero reflectance front surface and unit reflectance back surface \cite{tiedje1984}, $i.e.$,
\begin{eqnarray}
a(E) = 1 - e^{-2\alpha(E)L}
\end{eqnarray}
The factor of two in the above equation refers to the double pass of light because of the reflecting back surface. $\alpha(E)$ is the optical absorption coefficient, which is calculated as,
\begin{eqnarray}
\alpha(E) = \frac{2E}{\hbar c}{\sqrt\frac{\sqrt{(\epsilon_{1}(E))^{2} + (\epsilon_{2}(E))^{2}} - \epsilon_{1}(E)}{2}}
\end{eqnarray}

In Eq. (E17), second term is the dark current. The coefficient $J_{0}$ is the reverse saturation current, which can be calculated as the total (radiative and nonradiative) electron-hole recombination current at equilibrium in the dark.
\begin{eqnarray}
J_{0} = J^{r}_{0} + J^{nr}_{0} = \frac{J^{r}_{0}}{f_{r}}
\end{eqnarray}
In above equation, $f_{r}$ is known as the fraction of the radiative recombination current. In the SLME, $f_r$ is estimated using the following equation,
\begin{eqnarray}
f_{r} = e^{(E_{g} - E^{da}_{g}/kT)}
\end{eqnarray}
Here, $E_{g}$ is called the fundamental band-gap, while $E_{g}^{da}$ is the direct allowed band-gap of a compound.

According to the concept of detailed balance, the amount of absorption and emission via cell surfaces must equal at equilibrium in the dark. Therefore, $J^{r}_{0}$ can be easily obtained from the rate at which the front surface absorbs black-body photons from the surrounding thermal bath, $i.e.$,
\begin{eqnarray}
J^{r}_{0} = e\pi\int^{\infty}_{0}a(E)I_{bb}(E, T)dE
\end{eqnarray}
Where $I_{bb}$ denotes the black-body spectrum at temperature $T$. We can express both the solar spectrum $I_{sun}$ and black-body spectrum $I_{bb}$ in terms of photon flux. 

Eq. (E16) may be rewritten so as to maximise the power density,
\begin{eqnarray}
\eta = \frac{P_{max}}{P_{in}} = \frac{\max [(J_{sc} - J_{0}(e^{eV/kT} - 1))V]_{V}}{\int^{\infty}_{0}EI_{sun}(E)dE}
\end{eqnarray}
 
Thus, the material property-based parameters $\alpha(E)$, $f_{r}$, $T$, and $L$ are required to estimate the SLME.

\subsection{\label{sec:level2}Thermoelectric properties}

The semi-classical Boltzmann transport equations are used to obtain the thermoelectric (TE) coefficients that can be represented by the following equations \cite{yang2016},
\begin{figure}
\includegraphics[width=8cm, height=6cm]{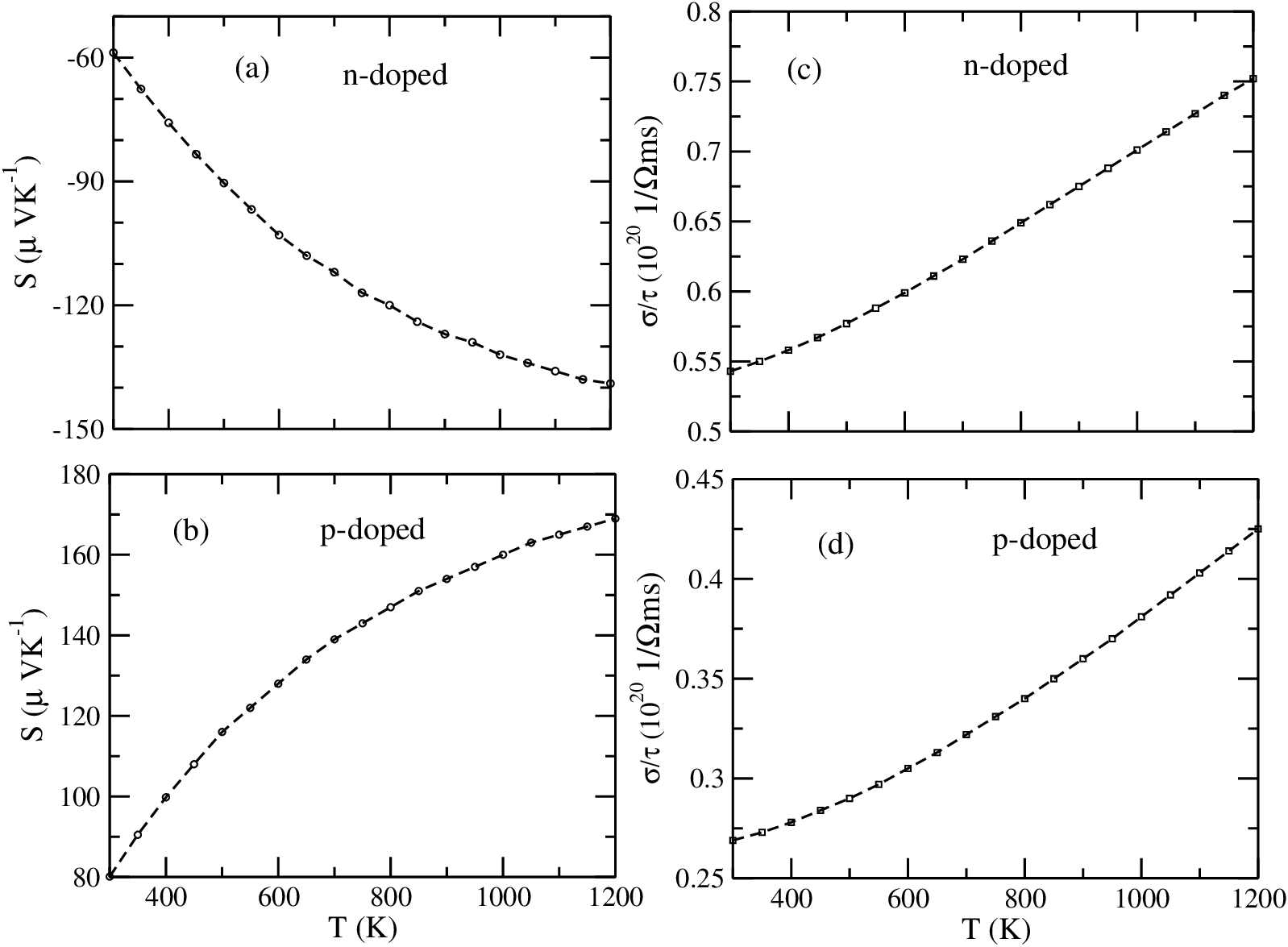} 
\caption{The calculated Seebeck coefficient ($S$) and electrical conductivity per relaxation time ($\sigma/\tau$) for $n$-type and $p$-type doped ScAgC.} 
\end{figure}

\begin{eqnarray}
S = \frac{e}{\sigma T} \int \sigma _{\alpha \beta }(\varepsilon) (\varepsilon - \mu) [\frac{-\partial f_{\mu}(T,\varepsilon)}{\partial \varepsilon }]  
\end{eqnarray}
and,
\begin{eqnarray}
\sigma _{\alpha \beta }(T, \mu) = \frac{1}{\Omega} \int \sigma _{\alpha \beta }(\varepsilon)[-\partial f_{\mu}(T,\varepsilon)] d\varepsilon   
\end{eqnarray}
Where, $\Omega, \mu, \sigma $, $S$, $T$, and $f$ represents the unit-cell volume, chemical potential, electrical conductivity, Seebeck coefficient, absolute temperature, and Fermi-Dirac distribution function, respectively. Also, $\alpha$ and $\beta $ are the tensor indices and $e$ is the charge carrier.

A good TE compound must have good electrical transport properties, which are measured by the power factor ($PF$), which can be written as \cite{yang2016},
\begin{eqnarray}
\textit{PF}= {S^{2}\sigma}
\end{eqnarray}

Also, good TE material exhibits a high dimension-less figure of merit ($ZT$) value, which is determined by,
\begin{eqnarray}
ZT= \frac{S^{2}\sigma T}{\kappa}
\end{eqnarray} 
The total thermal conductivity of a material is known as $\kappa$ = $\kappa_{e}$ + $\kappa_{ph}$, where $\kappa_{e}$ is the electrical thermal conductivity and $\kappa_{ph}$ is the lattice thermal conductivity.

\end{document}